# Engineering topological superconductors using surface atomic-layer/molecule hybrid materials

Takashi Uchihashi

International Centre for Materials Nanoarchitectonics (MANA), National Institute for Materials Science, 1-1, Namiki, Tsukuba, Ibaraki 305-0044, Japan

**Abstract**

Surface atomic-layer (SAL) superconductors consisting of epitaxially grown metal adatoms on a clean semiconductor surface have been recently established. Compared to conventional metal thin films, they have two important features: i) space-inversion symmetry breaking throughout the system and ii) high sensitivity to surface adsorption of foreign species. These potentially lead to manifestation of the Rashba effect and a Zeeman field exerted by adsorbed magnetic organic molecules. After introduction of archetypical SAL superconductor Si(111)-($\sqrt{7}\times\sqrt{3}$)-In, we describe how these features are utilized to engineer topological superconductor with Majorana fermions, and discuss its promises and expected challenges.

# 1. Introduction

The last decade has seen an explosive growth of research on topological materials, and among all, topological insulator has attracted immense attentions since its experimental finding [1-5]. It is induced by a strong spin-orbit coupling and is characterized by a full excitation energy gap within the bulk and gapless edge states at the boundary, which are protected by time-reversal symmetry. Equally very important is the study of topological superconductor and Majorana fermions as its zero-energy quasiparticle excitations [6, 7]. This exotic state of matter is not only fundamentally significant in terms of the non-Abelian statistics but also technologically relevant because manipulation of Majorana fermions will enable fault-tolerant quantum computing [8]. Nevertheless, experimental signatures reported so far has been very limited due to the technical difficulties [9-16], mostly concentrating on one-dimensional (1D) systems. One of the challenges toward this goal is the need of fabricating a hybrid system, which consists of different materials such as III-V semiconductor, metal-based superconductor, magnetic insulator, and topological insulator [17, 18]. Here, interfaces with different materials should play crucial roles, but it is generally difficult to fabricate atomically well-defined and high-quality interfaces of hybrid materials.

In this paper, we propose to engineer topological superconductors based on recently found surface atomic-layer (SAL) superconductors [19-22]. This class of materials consists of a long-range ordered layer of metal adatoms on the clean surface of a semiconductor such as silicon or germanium [23]. Because of chemical bonding between the surface adatoms and the substrate lattice, they form a unique atomic structure and electronic states distinct from the bulk counterparts. (Technically, they are called surface reconstructions.) Compared to conventional thin metal films on a semiconductor substrate, it possesses a high crystallinity and atomically clean and flat surfaces/interfaces, and hence can be regarded as an ideal two-dimensional (2D) electron system. Since such atomic-layer materials totally consist of surface, they are also unique in terms of the following aspects; i) space-inversion symmetry is broken, naturally leading to the Rashba effect due to the spin-orbit interaction [24], ii) the electronic states are sensitive to surface adsorption of foreign species. We propose that, when combined with a layer of magnetic organic molecules, these features could be used to realize a topological superconductor predicted by Sau et al. [17]. After introduction of the archetypical SAL superconductor Si(111)-($\sqrt{7}\times\sqrt{3}$)-In [referred to as ($\sqrt{7}\times\sqrt{3}$)-In], we describe the material design and discuss its promises and expected challenges.

# 2. Surface atomic-layer superconductor Si(111)-($\sqrt{7}\times\sqrt{3}$)-In

In this section, we introduce the ($\sqrt{7}\times\sqrt{3}$)-In surface as an archetypical SAL superconductor. This surface material consists of indium atomic layers epitaxially grown on a clean Si(111) surface [25]; here the indium atomic layer takes a commensurate registry against the silicon lattice with a $\sqrt{7}\times\sqrt{3}$ periodicity. Figures 1(a)(c) show the representative results of low energy electron diffraction (LEED) and scanning tunneling microscope (STM) measurements taken for this surface. Both LEED and STM images show the presence of well-defined √7×√3 periodicity. Regarding the electronic states, angle-resolved photoemission spectroscopy (ARPES) measurements revealed the presence of highly dispersed bands and a clear Fermi surface, which was modeled by an isotropic 2D Fermi circle folded at the Brillouin-zone boundary [26]. Despite these facts, the detailed atomic structure of this surface is still unknown. Early studies based on STM observation indicated that the surface layer consists of single atomic layer of indium [25], but later the existence of a double-layer-thick phase was also suggested [27]. Recent first-principle calculations proposed models where the indium layer has a double atomic-layer height [28, 29] [see figure 1(b) for the model proposed by Park and Kang], which successfully reproduced the result of ARPES. Nevertheless, they could not explain the observed STM image [figure 1(c)].

Consistent with the presence of a clear Fermi surface, electron transport measurements revealed a metallic conduction down to ~ 10K, i.e. dR/dT > 0 (R: sheet resistance, T: temperature) [30]. Note that conventional metal thin films typically exhibit insulating behaviors at low temperatures (i.e. dR/dT < 0), when the film thickness approaches the atomic-scale limit. This is due to the disorder-induced Anderson localization and the electron correlation effect [31]. The pristine (√7×√3)-In surface has no such localization tendency down to the lowest temperature. The result can be attributed to the well-ordered crystalline structure of the system, in clear contrast to disordered granular or amorphous films used in conventional experiments.

Recently, the (√7×√3)-In surface was found to undergo a superconducting transition at T ~3K against the prevailing belief that superconductivity cannot exist in an ideal 2D system. This was first clarified by observation of the superconducting energy gap using a low-temperature STM [19]. Subsequently, electron transport measurements demonstrated emergence of zero resistance over a macroscopic scale, which is a direct evidence of superconductivity [20-22]. Figure 2(a) shows a representative result of a four-terminal transport measurement on the (√7×√3)-In. The inset shows the optical microscope image of the actual sample and the configuration for the transport measurement. Following the gradual decrease in resistance that signifies the metallic behavior, the sample exhibits a sharp transition to a zero-resistance state as temperature is lowered below 3K. Temperature dependent current-voltage (I-V) characteristics also show a clear onset of superconductivity around 3K [figure 2(b)]. The inset of figure 2(b) shows superconducting critical current density ($J_c$) determined from the onset of the normal conduction. An example of STM observation of the superconducting (√7×√3)-In surface is displayed in figure 2(c), where the differential conductance (dI/dV) measured at zero bias voltage is mapped. The dashed line indicates the boundary of a (√7×√3)-In domain region. The two bright spots show the presence of vortices where superconductivity is strongly suppressed by application of magnetic field [32]. Representative dI/dV spectra obtained under magnetic field of 0.04 T are displayed in figure 2(d). The solid and dashed lines correspond to data taken outside of a vortex and at the center of a vortex, respectively. The former exhibits a clear superconducting energy gap (Δ = 0.39 meV) and coherence peaks, both of which are strongly suppressed in the latter as expected.

Apart from the (√7×√3)-In, similar surface systems of metal adatoms on silicon or germanium have been extensively studied [23]. Many of these materials have metallic electronic states judging from the presence of the Fermi surface, and some of them have already been confirmed to be metallic (Ge(111)-β(√3×√3)-Pb) [33] or superconducting (Si(111)-SIC-Pb, Si(111)-(√7×√3)-Pb) [19, 21] from transport and/or STM measurements at low temperatures. Furthermore, recently found one-unit-cell thick FeSe layers grown on a $SrTiO_3$ substrate may also be included in this category. Surprisingly, this material exhibits a very high transition temperature $T_c > 40$ K, presumably due to a strong interaction with the substrate [34, 35]. Naturally we can expect that more SAL superconductors will be discovered as the research is continued.

## 3. Surface-molecule hybrid system: route to topological superconductors

SAL superconductors possess unique properties that are distinct from bulk or conventional thin-film superconductors. In this section, we describe how they can be utilized to engineer a topological superconductor with Majorana fermions.

First, since it is located on the substrate surface, space-inversion symmetry is broken throughout the whole system regardless of its crystal structure. When combined with the spin-orbit interaction, this leads to the Rashba effect expressed by a Hamiltonian

$$H_{Rashba} = \alpha(\vec{\sigma} \times \vec{p}) \cdot \hat{z} \quad (1)$$

where $\alpha$ is the strength of Rashba spin-orbit coupling, $\vec{\sigma} = (\sigma_x, \sigma_y, \sigma_z)$ are the Pauli matrices, $\vec{p}$ is the momentum operator, and $\hat{z}$ is the unit vector in the surface normal direction [24]. Suppose free-electron-like electronic states with a parabolic energy dispersion in the 2D system. In the absence of the Rashba effect, the band is doubly degenerated with spin up and down states [see figure 3(a)]. This band becomes spin-split with introduction of the Rashba effect. The resulting two bands are chirally spin-polarized in opposite directions [figure 3(b)] [36]. Indeed, the Rashba effect and spin-polarized energy bands have already been observed in this class of surface atomic-layer materials, e.g. for Si(111)-β(√3×√3)-Bi [37], Si(111)-(1×1)-Tl [38], and Ge(111)-β(√3×√3)-Pb [39], although the actual electronic and spin structures are more complicated. The observed Rashba effect is strong, with $\alpha = 0.2 - 2.3$ eV·Å and Rashba energy scale $E_R (= m^*\alpha^2/2\hbar^2) = 20 - 120$ meV. Note that metals used in these surfaces can exhibit superconductivity in bulk (Tl, Pb) or amorphous film (Bi) forms [40]. It is therefore envisioned that SAL superconductors with large Rashba splitting can be realized.

Second, since the whole part of the atomic layer is exposed to the surface, its electronic states can be strongly modified by surface adsorption of foreign species. Particularly interesting is adsorption of organic molecules because they can self-assemble to form ordered structures on metallic surfaces without disrupting the substrate [41, 42]. Suppose that magnetic organic molecules, e.g. phthalocyanine molecules coordinated with magnetic atoms at center, are adsorbed on a metal surface. The local spins of magnetic molecules generally have an exchange interaction with the conduction electrons in the substrate (s-d interaction) with an exchange energy J ~1 eV [43-45]. Furthermore, they may become magnetically coupled through the substrate or the molecular orbitals [46]. For example, two-dimensional networks of metal-organic complex including 2,4,6-tris(4-pyridyl)-1,3,5-triazine (T4PT) or 7,7,8,8-tetracyanoquinodimethane (TCNQ) molecules were found to exhibit ferromagnetic coupling due to the intermolecular hybridization at low temperatures [47-49]. In this case, the effect of the ferromagnetically aligned local spins on the conduction electrons, if it exists, can be viewed as a Zeeman field (i.e. an effective magnetic field)

$$B_{eff} = \frac{nJ\langle S_z \rangle}{g\mu_B} \quad (2)$$

with a Hamiltonian

$$H_{Zeeman} = \mu_B B_{eff} \sigma_z \quad (3)$$

[45, 50]. Here *n* is the number of local spins per surface unit cell, $\langle S_z \rangle$ is the statistical average of the local spins, and we assume that the spins and Zeeman field $B_{eff}$ are aligned in the surface normal (z) direction. For a bulk metal substrate, however, the effect on the conduction electrons should be limited because the exchange interaction with the local spins applies only to their close neighbors. On the contrary, for a surface system considered here, the exchange interaction may strongly influence the electronic properties of the *total* system. This is because the conduction electrons are confined to the atomically thin layer in direct contact with the local spins.

Based on these unique features, we propose to fabricate a 2D hybrid material composed of a SAL superconductor and ferromagnetically coupled magnetic organic molecules [see figure 4(a)]. Here we have i) chirally spin-polarized electronic states due to the Rashba effect in the superconducting layer and ii) a Zeeman field due to the exchange interaction exerted by magnetic molecules. We assume that the superconductivity in the atomic layer is of conventional s-wave character. This setup is exactly a materialization of 2D topological superconductor predicted by Sau et al. and by related work [17, 51-53]. Their theory considers free-electron-like 2D electronic states with both Rashba and Zeeman spin-splittings. Without superconductivity, the Hamiltonian is written as

$$H_0 = H_{kinetic} + H_{Rashba} + H_{Zeeman}$$
$$= \frac{\vec{p}^2}{2m} - \epsilon_F + \alpha(\vec{\sigma} \times \vec{p}) \cdot \hat{z} + \mu_B B_z \sigma_z \quad (4)$$

where $\epsilon_F$ is the Fermi energy (chemical potential) measured from the bottom of the electron band in the absence of the Rashba spin-splitting. Introduction of superconductivity is described using the Bogoliubov-de Gennes equation that includes a finite superconducting energy gap function $\Delta$. The above Hamiltonian gives two chirally spin-polarized energy bands, the degeneracy of which is lifted at k = 0 due to opening of an energy gap $\mu_B B_z$. In this situation, when the Fermi level is located within the energy gap, the inner Fermi surface disappears, leaving only a single chirally spin-polarized Fermi surface [figure 3(c)]. In the presence of s-wave superconductivity, it acquires topological properties equivalent to the chiral p-wave ($p_x+ip_y$ wave) superconductivity. The requirement of fulfilling this condition is

$$(\mu_B B_z)^2 > \epsilon_F^2 + \Delta^2. \qquad (5)$$

An important consequence of this unconventional superconducting state is emergence of Majorana fermions within a vortex or at the edge of a domain region. This scenario has widely been accepted as one of the closest route to realization of topological superconductivity and Majorana fermions. Our surface-molecule hybrid superconductor may satisfy these conditions. For a system with multiple electronic bands, the same argument applies if one of the band edges is close to the Fermi level. In this case, for this relevant band, the Fermi level should be located within the energy gap $\mu_B B_z$ created by application of Zeeman field [see figure 3(d)]. Despite the presence of the other bands, the topological character of the system is retained since the number of spin-split Fermi surfaces is odd [9, 54, 55]. The "effective Fermi energy" $\epsilon_F'$ is then determined from this particular energy band and may take a small value. This enlarges the opportunities of realizing topological superconductor in a real material.

In the original proposal in Ref.[17], topological superconductivity emerges in a 2D Rashba semiconductor that is sandwiched between a s-wave superconductor and a magnetic insulator [see figure 4(b)]. Superconductivity and Zeeman field are introduced into the semiconductor layer by proximity effects. However, interfaces between different materials such as metal, semiconductor, and magnetic insulator are generally far from ideal. They typically suffer from considerable roughness and intermixing because of lattice mismatching, different character of bonding, and contamination [56]. Because of an imperfect interface, proximity-induced superconductivity is known to have a "soft gap" (i.e., the continuous subgap-states exist below $\Delta$), which destroys the topological properties [9, 57-60]. Our proposal based on atomic and molecular layers can circumvent this problem. There is no need for the superconducting proximity effect because the superconductor layer itself has Rashba spin-split states due to the space-inversion symmetry breaking. The Zeeman field still needs to be induced from the magnetic layer in the proximity, but its interface can be atomically clean and well-defined. The proposed configuration also has an advantage in terms of experimental detection. Since the superconducting layer is exposed to the surface through the molecular layer, it can be easily accessed using standard experimental techniques such as STM and ARPES. In contrast, it is not clear how to experimentally probe the electronic states and superconducting properties of the embedded semiconductor layer in figure 4(b).

The feasibility of fabricating a surface-molecule hybrid superconductor is shown in figures

4(c)-(e). In the experiment, Co-phthalocyanine (Pc), an archetypical magnetic molecule, was sublimed onto a (√7×√3)-In surface to form single and double layers of molecules [see the inset of figure 4(c) for the chemical structure of CoPc]. STM observations found that the first layer of CoPc forms a deformed triangular lattice, reflecting the symmetry of the Si(111) substrate and the (√7×√3)-In surface [figure 4(c)(d)]. This suggests a relatively strong interaction with the molecules and the indium layer because Pc molecules generally form a square lattice on a weekly interacting substrate due to its four-fold symmetric structure [46, 61, 62]. Nevertheless, our preliminary electron transport measurements have confirmed that the sample becomes superconducting at $T_c$ nearly equal to that of the pristine (√7×√3)-In. In contrast, the second CoPc layer forms a square lattice indicating a weak interaction with the underlying layer [figure 4(e)]. The spins of CoPc molecular are likely to be quenched for the first layer but to survive for the second layer, in the light of an experiment on a similar CoPc system [61]. The magnetic properties of surface molecules can be detected using inelastic-electron-tunneling-spectroscopy (IETS) STM [61, 63, 64], spin-polarized (SP) STM [49, 65, 66], and x-ray magnetic circular dichroism (XMCD) [47, 48, 67, 68]. The presence and the effect of molecular spins on a superconductor and formation of in-gap bound states (Shiba states) were also studied using STM [62]. Such measurements could be applied to a variety of surface/molecule system to study hybrid superconductors considered here.

Here we discuss expected problems in creating topological superconductor based on our proposal. First, superconductivity is generally suppressed by a Zeeman field because of the Pauli pair breaking effect. (Note that the orbital pair breaking effect can be neglected here since the stray magnetic field from the molecular layer is weak.) This might suggest that the coexistence of the magnetic molecular layer and the atomic-layer superconductor is unlikely. However, the superconductor considered here has spin-split Fermi surfaces due to the Rashba effect. In this case, the effect of the Zeeman field on superconductivity is negligibly small [69]. Indeed, studies on bulk heavy fermion superconductors have shown that superconductivity can be compatible with strong magnetic (Zeeman) field when the electronic states are Rashba spin-split [70]. Regarding the 2D superconductor, amorphous Pb monolayers grown on clean GaAs substrates were found to be surprisingly robust against a parallel magnetic field, which was attributed to the Rashba spin-splitting [71].

A more serious challenge will be the tuning of the Fermi level of the system to fulfill the condition of equation (5). To realize topological superconductivity, it is essential to adjust the Fermi level within the energy gap $\mu_B B_z$ [see figure 3(c)]; otherwise, two chirally spin-polarized Fermi surfaces survive to cancel their topological properties. This could be a serious problem because it is generally difficult to adjust the Fermi level of a metal through carrier doping. However, since the thickness of a surface atomic layer is comparable to the typical screening length in metals, tuning of the Fermi level through carrier doping from the surface should be possible. The molecular layer, which plays the role of the source of Zeeman field, can also donate charges into the underlying substrate [49, 72] and hence can help tune the Fermi level. We note that some of surface materials have a small and controllable Fermi energy. For example, the Si(111)-(√3×√3)-Ag surface has a nearly parabolic conduction band, and its Fermi energy $\epsilon_F$ can be controlled between 0 eV $< \epsilon_F <$ 0.8 eV through charge transfer from dopant adatoms [73]. Such a control should be possible if the surface layer is of semiconducting character, i.e. the Fermi level is located within or near the energy gap [1]. Very recently, introduction of superlattice (periodic potential modulation) on the superconducting 2D layer using organic molecules was proposed to open an energy gap near the Fermi level and to effectively reduce the Fermi energy [74]. If the number of the resultant spin-split Fermi surfaces is odd, it leads to manifestation of topological superconductivity as discussed above for the multiple electronic bands. While STM manipulation of individual molecules was considered to create a superlattice [74, 75], molecular self-assembly will also be useful since the resulting lattice structure may be flexibly controlled through rational design of molecules [42].

Relating to the Fermi-level tuning problem, a small energy gap $\mu_B B_z$ created by Zeeman field could be another obstacle. The Zeeman energy exerted by exchange interaction can be estimated from equation (2). If we take CoPc on the (√7×√3)-In surface for example, n = (areal density of CoPc)/(areal density of In atoms) = 0.05. Assuming J= 1 eV, $\langle S_z \rangle = 1/2$, and g = 2, Zeeman energy is estimated to be $\mu_B B_{eff} \sim 0.01$ eV. This is not large enough to satisfy the condition of equation (5) if $\epsilon_F$ is taken to be the typical Fermi energy of metals. Hence, it is essential to tune the Fermi level so that $\epsilon_F$ is reduced to a comparable energy scale. Fortunately, the requirement is less stringent for the multiple bands because the "effective Fermi energy" $\epsilon'_F$ may be reduced considerably as discussed above. In terms of the magnetic layer, the Zeeman energy can be enhanced by increasing the areal density of local spins $n\langle S_z \rangle$ and/or exchange interaction J. While $n\langle S_z \rangle$ can be increased by choosing a molecule with smaller dimensions and larger spins, J is strengthened by a stronger electronic coupling between the molecule and the conduction electron. Here also rational design of appropriate molecules will be a key to success.

Finally, we briefly mention how to identify the topological superconductivity through detection of Majorana fermions. Theoretically, it is well established that a Majorana fermion emerges within a vortex created in a 2D topological superconductor with chiral p-wave characteristics [6, 7, 17, 18, 52]. Scanning tunneling spectroscopy observation of a zero-bias peak with a quantized conductance of $2e^2/h$ at the vortex center will be a direct evidence of a Majorana fermion, since it reflects the electron-hole neutrality of a Majorana fermion through the resonant Andreev reflection [11, 51, 76]. We note that a vortex in a conventional superconductor can also have a zero-bias peak because of quasiparticle interference within a pair potential barrier [11, 77]. Since the amplitude of the zero-bias peak is strongly dependent of electron coherence within the system, the experiment should be performed using a high-quality sample and at very low temperatures. Strong spin-sensitivity of the Majorana-Fermion induced resonant Andreev reflection [78] may be investigated using a SP-STM to give a conclusive result. Alternatively, we can resort to observation of an Majorana fermion at the edge [76]. If the topological superconducting layer is confined within a domain region, the bulk-edge correspondence ensures that edge states emerge, which can have a Majorana fermion as a zero-energy bound state. This edge Majorana fermion exists only when there are an odd-number of vortices within the confined region [53, 79, 80]. This is a natural consequence from the fact that Majorana fermions always appear as a pair because they are halves of a conventional fermion. Their existence and the dependence on the parity of the vortex number can be detected through STM observation of the zero-bias peak. Such an experiment is conceivable as a natural extension of the one shown in figure 2(c), where the (√7×√3)-In region enclosed by the dashed line should be replaced by a 2D topological superconductor. Since our proposed hybrid superconductor is exposed to the surface through a molecular layer, access to the Majorana fermions within a vortex or at the edge will be straightforward in an STM experiment.

---

[1] In terms of emergence of superconductivity, a "doped semiconductor" with a small Fermi energy $\epsilon_F$ is disadvantageous in a 3D system, because $T_c$ is exponentially dependent on the density of state at the Fermi level $\rho(\epsilon_F)$ in the BCS theory, where $\rho(\epsilon_F) \propto \epsilon_F^{1/2}$ for a free-electron-like 3D system. However, since the surface atomic layer is a strictly 2D system, $\rho(\epsilon_F)$ is independent of $\epsilon_F$ in an ideal case. Hence a small Fermi energy is not disadvantageous in this respect.

## 4. Conclusion

We have described how a hybrid-type topological superconductor may be materialized using a SAL superconductor and magnetic organic molecules. The advantages of our proposal are i) having a high-quality and atomically defined surface and interface for a hybrid material, ii) no need of proximity effect for inducing superconductivity into a Rashba system, and iii) readily experimental access to the superconducting layer, e.g. using an STM. Particularly, feature iii) and flexibility of material choice are advantages over other hybrid systems based on 2D superconductivity at heterostructures [81, 82]. On the other hand, one of the challenges is to adjust the Fermi level so that the (effective) Fermi energy $\epsilon_F$ can be smaller than the Zeeman energy to fulfill the requirement $(\mu_B B_z)^2 > \epsilon_F^2 + \Delta^2$. Investigation into the detailed electronic and spin structure of SAL superconductors using, e. g. ARPES and first principle calculations will be necessary toward this goal. Rational design of magnetic organic molecules to realize a ferromagnetically coupled layer causing a large Zeeman field will also be crucial to fulfill the above criteria.

The author thanks S. Yoshizawa, P. Mishra, M. Yamamoto, H. Kim, and Y. Hasegawa for obtaining data presented here. He also thanks X. Hu, T. Kawakami, and T. Okamoto for fruitful discussions on topological materials and non-centrosymmetric superconductors. M. Aono is acknowledged for his advice and encouragement. This work was financially supported by JSPS under KAKENHI Grants No. 25247053 /26610107 and by World Premier International Research Center (WPI) Initiative on Materials Nanoarchitectonics, MEXT, Japan.

**Figure captions**

**Figure 1.** Atomic structure characterization of the (√7×√3)-In surface. (a) Representative LEED pattern (beam energy E = 86 eV) [20]. The dashed lines indicate the reciprocal unit cell. (b) Double-layer indium structure model of the (√7×√3)-In surface (blue spheres: In, yellow spheres: Si). Reprinted figure with permission from Park J and Kang M 2012 *Phys. Rev. Lett.* **109** 166102. Copyright (2012) by the American Physical Society. (c) STM image of a (√7×√3)-In surface (Sample bias voltage V = 400 mV, tunneling current I = 121 pA). The dashed lines indicate the unit cell in the real space.

**Figure 2.** Superconducting properties of the (√7×√3)-In surface. (a) Temperature dependence of the sheet resistance (2D conductivity) of a (√7×√3)-In surface (bias current I = 1 μA) [22]. The inset shows the actual sample and a schematic illustration of the four-terminal measurement configuration. The dashed rectangle shows the target area for transport measurement. (b) Temperature dependence of I-V characteristics (1.77 K< T < 3.11 K) [20]. The bias current was swept in the increasing direction. The inset shows the temperature dependence of critical current $I_c$ and 2D critical current density $J_{2D,c}$. (c) Zero-bias conductance image of a (√7×√3)-In surface taken using a low-temperature STM. The dashed line indicates that topographical boundary which separate the ordered (inside) and disordered (outside) regions of the (√7×√3)-In. Superconducting energy gap was observed only inside the boundary. The two bright features correspond to superconducting vortices created by magnetic field B = 0.04T. (d) dI/dV spectra taken on a (√7×√3)-In surface. The solid and dashed lines show spectra taken outside of and at the center of a vortex, respectively [32].

**Figure 3.** Schematic illustration of the influence of the Rashba and Zeeman effects on the energy band structure of a free-electron-like 2D electron system. The Fermi surfaces are indicated by solid lines, which are determined from the cross section between the energy bands and the $E = \epsilon_F$ planes. The arrows show the spin polarization at individual Fermi surfaces. (a)-(c): single band, (d): multiple bands. (a) Without the Rashba and Zeeman effects. (b) Only with the Rashba effect. (c)(d) With the Rashba and Zeeman effects. The Fermi level is adjusted within the energy gap created by the Zeeman effect, leading to manifestation of topological superconductivity of effective chiral p-wave.

**Figure 4.** Design of surface-molecule hybrid materials exhibiting topological superconductivity. (a) Schematic drawing of the hybrid structure consisting of magnetic molecules assembled on a surface atomic-layer superconductor. The two layers are supported on an insulating substrate. (b) Schematic drawing of the original proposal by Sau et al [17, 51], consisting of (s-wave) superconductor, Rashba semiconductor, and magnetic insulator. The arrows in (a) and (b) indicate ferromagnetically aligned

local spins. (c) Topographic STM image of a single layer of CoPc molecules assembled on a (√7×√3)-In surface. The inset shows chemical structure of a CoPc molecule (blue sphere: H, black spheres: C, red spheres: N, green spheres: Co). (d)(e) Magnified STM image of a (d) single and (e) double layer of CoPc molecules on a (√7×√3)-In surface. In (d) the molecules are aligned in a distorted triangular lattice while a square lattice is visible in (e).

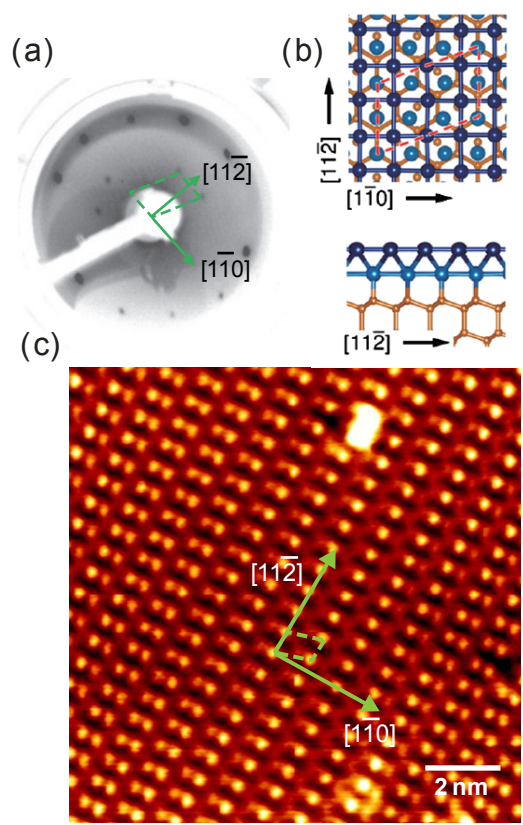

Figure 1

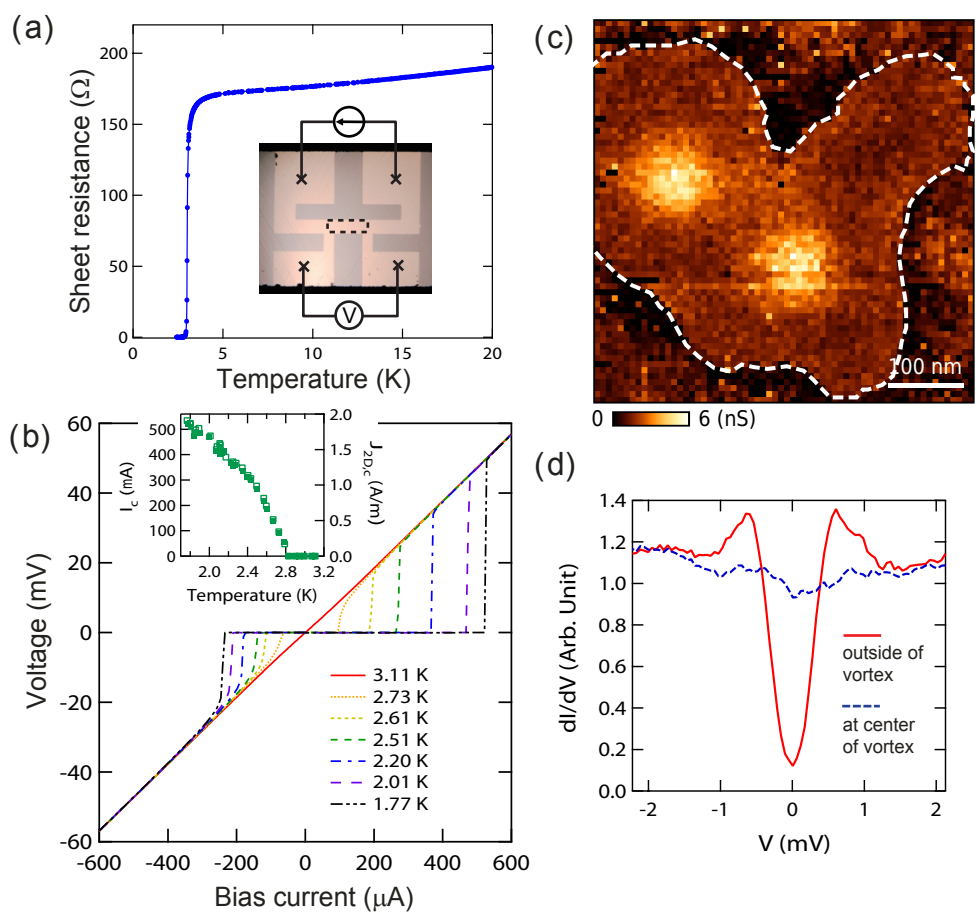

Figure 2

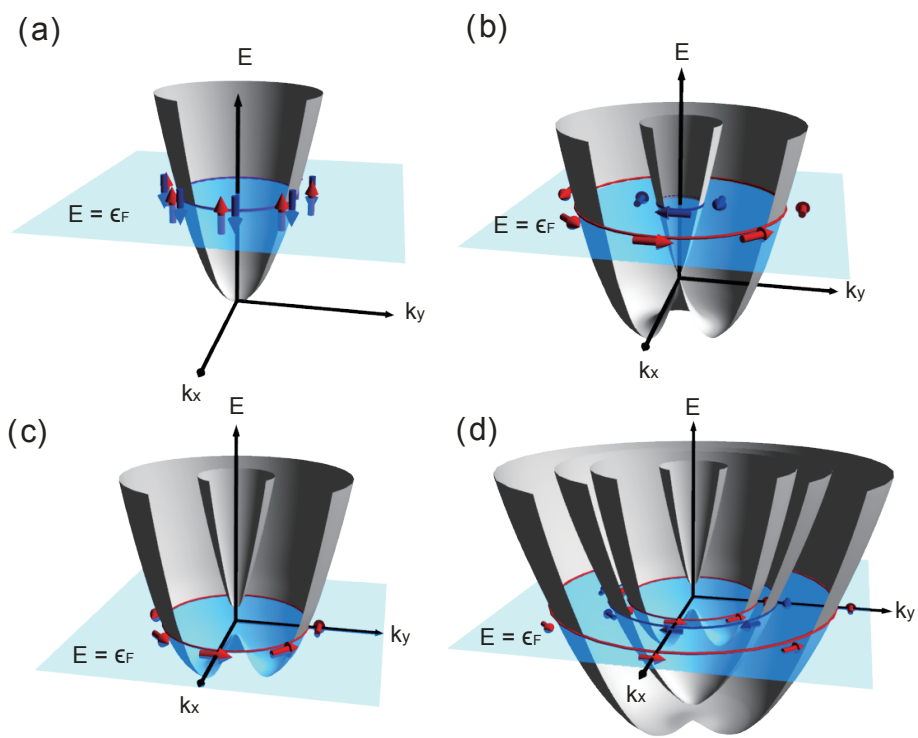

Figure 3

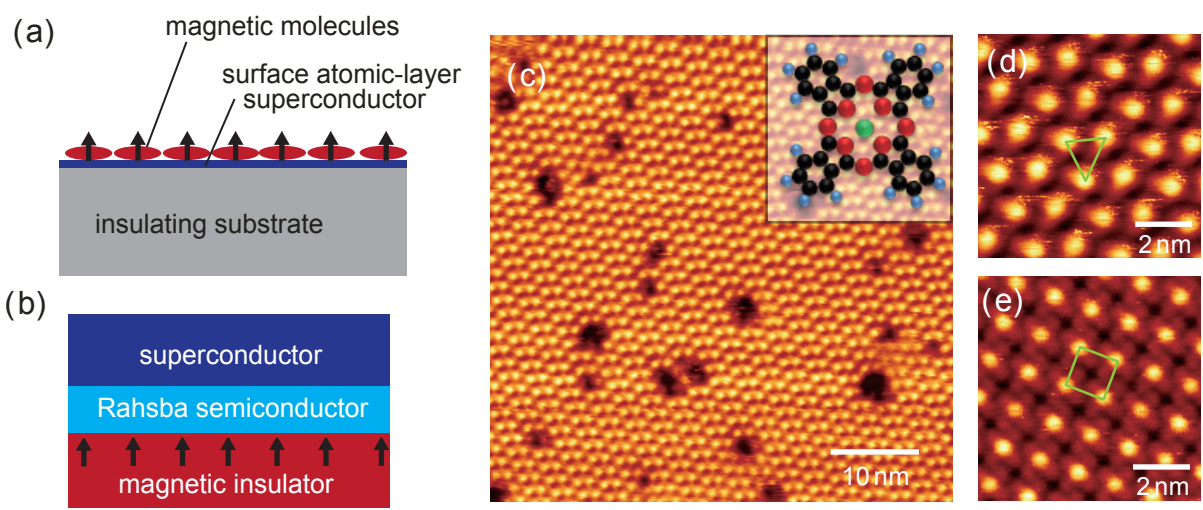

Figure 4